# How Community Agreements Can Improve Workplace Culture in Physics

Emanuela Barzi*

*Fermi National Accelerator Laboratory, Batavia, IL 60510, USA, and Ohio State University, Columbus, OH 43210, USA*

Simonetta Liuti

*University of Virginia, Charlottesville, VA 22904, USA*

Christine Nattrass

*University of Tennessee, Knoxville, TN 37996, USA*

Roxanne Springer

*Duke University, Durham, NC 27708, USA*

Charles H. Bennett

*IBM T.J. Watson Research Center, Yorktown Heights, NY 10598, USA*

CO SIGNATORIES:

Barry C. Barish

*California Institute of Technology, Pasadena, CA 91125, USA, and U.C. Riverside, Riverside, CA 92521, USA*

Jonathan A. Bagger

*Johns Hopkins University, Baltimore, MD 21218, USA*

Robert Rosner

*The University of Chicago, Chicago, IL 60637, USA*

Michael P. Marder

*The University of Texas at Austin, Austin, TX 78712, USA*

Sergio Bertolucci

*University of Bologna, Bologna 40126, Italy*

Michael S. Witherell

*UC Berkeley, Berkeley, CA 94720, USA*

JoAnne Hewett, Hirohisa A. Tanaka, Nan Phinney

*SLAC National Accelerator Laboratory, Stanford University, Menlo Park, CA 94025, USA*

Haiyan Gao

*Brookhaven National Laboratory, Upton, NY 11973-5000, USA, and Duke University, Durham, NC 27708, USA*

Michiko Minty, Jerome Lauret, Elizabeth Worcester

*Brookhaven National Laboratory, Upton, NY 11973-5000, USA*

Ramona Vogt

*Lawrence Livermore National Laboratory, Livermore, CA 94551, USA and University of California at Davis, Davis, CA 95616, USA*

William Collins, Spencer Klein

*Lawrence Berkeley National Laboratory and University of California, Berkeley, CA 94720, USA*

Ernst Sichtermann, Andrew Doran, John Arrington, Reiner Kruecken, Mateusz Ploskon

*Lawrence Berkeley National Laboratory, Berkeley, CA 94720, USA*

Sam Bader

*Argonne National Laboratory, Lemont, IL 60439, USA*

Alan Hurd, Sowjanya Gollapinni

*Los Alamos National Laboratory, Los Alamos, NM 87545, USA*

Robert Roser
*Idaho National Lab, Idaho Falls, ID 83415, USA*

Alexander V. Zlobin, Teri Dykhuis, Michelle A. Ibrahim, Sudeshna Ganguly, Roshanda Spillers-Nowlin, Anne Heavey, Lauren Hsu, Dylan Temples, Sho Uemura, Kavin Ammigan, Claire Lee, Nicholas Bornman, Finley Novak, Anna Mazzacane, Christine A. Ader, Elizabeth Sexton-Kennedy, Jessica Esquivel
*Fermi National Accelerator Laboratory, Batavia, IL 60510, USA*

Narbe Kalantarians
*Virginia Union University, Richmond, VA 23220, USA, and Jefferson Lab, Newport News, VA 23606, USA*

Jacquelyn Noronha-Hostler, Jorge Noronha, Vidushi Adlakha
*University of Illinois at Urbana-Champaign, Urbana, IL 61801, USA*

Timothy A. Raines
*Indiana Institute of Technology, Fort Wayne, IN 46803, USA*

Meredith D. Betterton, Heather Lewandowski
*University of Colorado Boulder, Boulder, CO 80309, USA*

Beverly K. Hartline
*Montana Technological University, Butte, MT 59701, USA*

Baha Balantekin
*University of Wisconsin, Madison, WI 53706, USA*

Senta Vicki Greene
*Vanderbilt University, Nashville, TN 37235, USA*

Sherry Yennello
*Texas A&M University, College Station, TX 77845, USA*

Warren F. Rogers
*Indiana Wesleyan University, Marion, IN 46953, USA*

Laura Fields, Rebecca Surman
*University of Notre Dame, Notre Dame, IN 46556, USA*

Rosi Reed
*Lehigh University, Bethlehem, PA 18015, USA*

Helen Caines
*Yale University, New Haven, CT 06518, USA*

Vianney Gimenez-Pinto
*Lincoln University of Missouri, Jefferson City, MO 65101, USA*

John Lajoie
*Iowa State University, Ames, IA 50011, USA*

Marie Boer
*Virginia Tech, Blacksburg, VA 24060, USA*

Sevil Salur
*Rutgers, The State University of NJ, Piscataway, NJ 08854, USA*

Amy Bug Graves
*Swarthmore College, Swarthmore, PA 19081, USA*

Meenakshi Narain
*Brown University, Providence, RI 02912, USA*

Saptaparna Bhattacharya
*Northwestern University, Evanston, IL 60208, USA*

Nora Berrah
*University of Connecticut, Storrs, CT 06269, USA*
Michael Troxel
*Duke University, Durham, NC 27708-0305, USA*
Ágnes Mócsy
*Facility for Rare Isotope Beams, Michigan State University, MI 48824, USA and Pratt Institute, NY 11205, USA*
Michael Thoennessen, F.M. Nunes, R.G.T. Zegers, Dean Lee
*Facility for Rare Isotope Beams, Michigan State University, MI 48824, USA*
Ayres Freitas
*PITT PACC, University of Pittsburgh, Pittsburgh, PA 15260, USA*
John Hauptman
*Iowa State University, Ames, IA 50011, USA*
Olga Evdokimov
*University of Illinois at Chicago, Chicago, IL 60607, USA*
Claudia Ratti
*University of Houston, Houston, TX 77204, USA*
Jennifer L. Klay
*California Polytechnic State University, San Luis Obispo, CA 93407, USA*
Maajida Murdock
*Morgan State University, Baltimore, MD 21251, USA*
Sanjay Reddy
*University of Washington, Seattle, WA 98195, USA*
Susan K. Blessing, Vandana Tripathi
*Florida State University, Tallahassee, FL 32301, USA*
Leo Hollberg
*Stanford University, Stanford, CA 94305, USA*
Akaa Daniell Ayangeakaa
*University of North Carolina at Chapel Hill, Chapel Hill, NC 29208, USA*
Giorgio Bellettini, Simone Donati
*University of Pisa, Pisa 56127, Italy*
Patrizia Azzi
*INFN, Section of Padova, Padova 35100, Italy*
Matthew G. Hannah
*University of Bayreuth, 95440 Bayreuth, Germany*
Juergen Reuter
*Deutsches Elektronen-Synchrotron DESY, 22607 Hamburg, Germany*
Carlota Andres
*CPHT, CNRS, Ecole polytechnique, IP Paris, F-91128 Palaiseau, France*
Frank Krauss
*IPPP Durham, Durham University, Durham DH1 3LE, United Kingdom*
Maria Longobardi
*University of Basel, 4056 Basel, Switzerland*
Ursula Keller
*ETH Zurich, Switzerland*

Elias Métral

*CERN, 1211 Geneva 23, Switzerland*

M. A. Mahmoud

*Center for High Energy Physics (CHEP-FU), Fayoum University, Egypt*

Akihiro Kikuchi

*National Institute for Materials Science, Tsukuba, Ibaraki 305-0047, Japan*

*barzi@fnal.gov

ABSTRACT

Equity, Diversity, and Inclusion (EDI) committees and Codes of Conduct (CoC) have become common in laboratories and physics departments across the country. However, very often these EDI committees and CoC are not equipped to provide practical consequences for violations, and therefore are mostly performative in nature. A considerable effort has been devoted by various groups within APS units and beyond the APS in developing instead what are now called Community Guidelines. Community Guidelines help implement the core principles in CoC, by setting expectations for participation in in-person events and virtual communication. When further accompanied by accountability and enforcement processes, they develop into Community Agreements. This White Paper discusses the elements necessary to create and implement an effective Community Agreement, reviews examples of Community Agreements in physics, and argues that physics collaborations, physics departments, and ultimately as many physics organizations as possible, however large or small, should have a Community Agreement in place. We advocate that Community Agreements should become part of the bylaws of any entity that has bylaws.

## 1. EXECUTIVE SUMMARY AND RECOMMENDATIONS

### 1.1. Summary

Community Agreements (CA) are the more modern term for what have been called "Codes of Conduct" (CoC): documents where a community outlines expectations for how community members interact in supportive, respectful, ethical, safe, welcoming, and inclusive ways. CoC that apply within universities, laboratories, and other physics organizations often use the same language as that of state or federal judicial systems such as Title IX. At the same time, they often lack implementation mechanisms that are adequate for a physics community. The result can often seem dispirited, slow, or ineffective, lacking transparency and accountability, as in "performative" activism.

The goal of a formal CA is to help maintain or achieve a good organizational climate, where every member of the organization perceives that ethical conduct will enhance their standing and reputation, leading to success to which all members have contributed and for which all can take credit. Conversely, in a bad or "toxic" organizational climate, members perceive that the only way to get or stay ahead is to disrespect their colleagues in small and large ways. Normal and salutary competition is replaced by a perceived need to flatter one's superiors, bully one's subordinates, and cover one's backside. The spectrum of bad behavior ranges from minor incivilities that break no explicit rule to blatant violations of professional integrity, such as submitting a scientific paper containing results that one or more of the authors believe to be false or plagiarized, or blunt discrimination of the kind prohibited by Title IX and its modern extensions.

To implement ethical procedures, it is necessary to understand the mechanisms by which discrimination and other kinds of unethical behavior occur. The most prevalent forms of discrimination are sexual harassment and retaliation. And since

2008, the Equal Employment Opportunity Commission (EEOC) has reported that retaliation is the most common discrimination finding in federal sector cases.

Key findings of the National Academy of Sciences (NAS) and the EEOC to improve workplace climate include:

1. Addressing the most common form of sexual harassment: gender harassment;

2. Moving beyond legal compliance to address culture and climate;

3. Improving transparency and accountability;

4. Diffusing the hierarchical and dependent relationship between employees and managers;

5. Providing support for the victim;

6. Making the entire community responsible for reducing and preventing harassment, discrimination and other forms of unethical behavior.

**1.2. Recommendations**

- All physics entities should have a Community Agreement (CA) in place, and abiding by the CA should be an explicit condition of membership in that entity. For entities with existing bylaws, CA could become part of those bylaws and violations of the CA should be taken at least as seriously as violations of other aspects of an entity's bylaws. CAs should include a list of well-defined expectations, a team of individuals tasked with accepting reports, as well as an enforcement mechanism that uses equitable standards of fact-finding, clear-cut consequences, and adjudication that is unbiased, as transparent as possible, and equal for all. The CA should be a living document, updated as needed to become increasingly fair and effective, but not so easy to change that implementation begins to appear arbitrary, or time-dependent.

- Depending on the nature and severity of the alleged unethical behavior, investigation procedures should balance transparency and privacy, while allowing cases to be adjudicated and appropriate sanctions (ranging from no action to expulsion from the organization) to be applied expeditiously. Criminal acts should be referred to legal authorities, while CAs typically address cultural issues. Complementary to their inability to exact civil or criminal penalties, membership organizations such as the APS have broad powers to vote to suspend or expel a member deemed to have damaged the reputation of the organization, without needing to prove that the member did anything illegal. CAs have to enable the community to punish behavior that, while not illegal, is detrimental to the health of the community.

- The tradeoff between transparency and confidentiality needs to be navigated while keeping in mind that not enough confidentiality early on discourages frank reporting and investigation, while too little transparency at the end leaves the community wondering if complaints have been properly investigated or even whether investigators are corruptly excusing friends and punishing enemies. In the intermediate term appropriately anonymized or redacted data can be made publicly available, as done in [17]. To maximize transparency, the rules for who can examine still-confidential data, and when, should be publicly known and hard to change.

- Organizations have sometimes hesitated to impose severe punishments, such as revocation of previously awarded honors or expulsion from the organization, for fear of civil lawsuits, especially when the finding of misconduct is based on confidential data. However, with the means to expeditiously and safely punish severe misconduct, the organization would be better able to avoid lawsuits of the opposite kind, by complainants who felt their complaints had been ignored and their rights violated by an organization too fearful of being sued for overzealous enforcement.

- All members of the physics entity should undergo regular interactive, role-playing training to obtain a deeper understanding of how to create a safe, welcoming, and inclusive community, as well as avoiding gray areas of professional conduct such as careless attribution and real or perceived conflicts of interest. This involves physicists reflecting on their own actions and by learning to constructively intervene in real time when witnessing unethical behavior by others.

- Each physics entity should encourage feedback from members about any and all interactions that may be hampering the effort to create a welcoming and inclusive culture. Minor missteps, bullying, talking over others, etc., are part of the ``incivility" in a culture that is the gateway to worse behavior. The CA should be nimble and nuanced enough to provide constructive and appropriate responses that will allow the community to heal and grow.

- Institutions can take concrete steps to reduce discrimination by making systemic changes that demonstrate how seriously they take this issue and that reflect that they are listening to those who speak up. This is in contrast with the policies and procedures that only protect the liability of the institution but are not effective in preventing misconduct. Hierarchical power structures, where power is concentrated in single individuals, are more likely to foster and sustain harassment, bullying and retaliation, while jeopardizing their members' ability to produce good science.

- Climate is a measurable quantity. To assess climate, the National Academies recommend doing a third-party anonymous survey, since waiting for complaints is an unreliable way of measuring climate in any scientific entity. If the climate is assessed as "poor," leadership should recognize the problem quickly and move beyond legal compliance to address its culture. If needed, restructuring HR and Legal Counsel offices could be considered. When hosts of collaborations, labs should also provide resources for development of their Community Agreements and offer full support in enforcing them. A healthier climate would also mitigate risk of lawsuits.

## 2. INTRODUCTION

Both diversity and inclusion find root in the larger ethical concept of equity, which is the fair treatment, access, opportunity, and advancement for all people, based on their actual talents and skills. It is reasonable to assume that the happiest and most productive societies are those based on equity and social justice, independent of the moral argument for creating equitable societies. Improving equity involves increasing justice and fairness within the procedures and processes of institutions or systems, as well as in their distribution of resources [1, 2]. The importance of this effort has been recognized by the Department of Energy's Office of Science, the largest federal sponsor of basic research in the physical sciences and a steward of public funding, which has introduced a new requirement to its solicitation processes: applicants must now submit a plan for Promoting Inclusive and Equitable Research, or PIER Plan, along with their research proposals [3].

Scientific organizations' and publications' Codes of Conduct (CoC) generally include scientists' duty to behave ethically, respectfully, and inclusively toward one another and to reveal potential conflicts of interest. Together with Equity, Diversity, and Inclusion (EDI) committees, Statements of Values and CoC have become ubiquitous in the past few years, in laboratories and physics departments alike across the country. This is due to the regrettable fact that equity in science is still a mirage, as shown by survey after survey [4]-[7]. The sharp contrast between the often-toxic environments found in scientific labs and in academia and healthier ones that are sometimes found in industries that are more customer oriented, for instance, has made the human culture in science a serious concern for all, and especially for the funding agencies. Things are better in some private sectors, thanks to the increase of research funding within Industrial-Organizational Psychology (I-OP), to mitigate the financial and reputation damages from the #MeToo and similar movements [1].

In 2002, psychologists identified personality traits, including narcissism, that undermine the work and well-being of others in a lasting and often irreversible way. Another, more recent, main finding that has garnered global interest is that executive leaders tend to be higher than average in narcissism — a personality trait characterized by an inflated sense of self-importance, a strong desire for power, and a propensity for manipulative behavior. Several social studies have shown that narcissism affects more than 18 percent of individuals in positions of high responsibility, CEOs for example, as compared to the 5 percent average in the general population [8, 9]. Narcissists often pursue and are selected for leadership positions. At the same time, they act in their own best interest, putting the needs and interests of the organization and others at risk. Research has shown that self-centeredness also hinders the flow of communication throughout the organization and impoverishes the climate. Unfortunately, statistics and studies speak clearly on how prevalent these negative traits are in physics.

This often-toxic climate is deleterious in general for scientific productivity in the country, and is remarkably antithetic to a modern society, where younger generations care a great deal about emotional intelligence and empathy [10]. Bearing witness to the keen sense of urgency in the community for a more just and livable workplace has become of critical importance to sustain the scientific knowledge and effort for future generations.

## 3. THE NEED FOR COMMUNITY AGREEMENTS

It is apparent that an equitable and just community does not spontaneously occur without effort. It will take sustained commitment, attention, action, measurement, reflection, and adjustment to achieve a welcoming and inclusive physics community. There are those who argue that equity and inclusion will naturally occur once a sufficient number of participants are those of previously marginalized communities, but the case of medicine, academic humanities, Ph.D.s in chemistry, etc., show that raw numbers are not sufficient. Instead, policy and culture change is required. We propose that Community Agreements are part of the solution.

The original establishment of Codes of Conduct (CoC) produced a number of unforeseen negative consequences. CoC that apply within universities, laboratories, and other physics organizations often use the same standards as state or federal judicial systems such as Title IX, which are inadequate to change the culture in physics [5]. As clearly explained in [5], to be effective organizations have to move beyond legal compliance, and improve transparency and accountability. On the other hand, CoC for collaborations are often written in such a way that the process of dealing with violations/imposing sanctions is either not defined or unenforceable. This is bound to lead to arbitrariness, and to generate mistrust in the process. The equivalent of these phenomena in an institution, i.e., when institutions do not have the means or neglect to enforce disciplinary sanctions as necessary, is called "institutional betrayal," because it causes additional harm to the victims [11].

To effect positive change, an EDI organizational initiative has to have clear and well-defined goals, as well as an enforcement policy that includes equitable standards of fact-finding, clear-cut consequences, and adjudication that is unbiased and equal for all.

All too often, one sees instead "performative activism," which is defined as "the act of advocating for a cause or issue to gain attention, support, or profit rather than caring about making a difference in the cause" [12, 13]. Another unintended deleterious consequence of CoC is their weaponization against the very people they were designed to protect. This occurs especially within rigid academic and bureaucratic organizational structures. For example, junior physicists complaining of abuse, or those belonging to underrepresented groups, can be accused of "uncollegial" language when the language reflects inexperience rather than malice, or worse, would have been tolerated if used by a person in a position of authority. Ref. [14] details the ways in which retaliation can thrive in organizations that have insufficient oversight and allow leaders to abuse power: "Organizations that foster a climate of aggression and bullying are more likely to have managers who abuse power and retaliate when claims are made. Other organizational factors that influence retaliation are: a lack of administrative policies discouraging retaliation; an authoritarian management culture; overly hierarchical organizations, where rank or organizational level is prized; high levels of task-related conflicts; reward systems and structures that promote competition; and the ability to isolate the accuser."

In order to implement ethical procedures, it is necessary to understand the mechanisms by which discrimination occurs. The most prevalent form of discrimination is sexual harassment and retaliation. As is well-known, the much needed effort in closing the "gender gap" in science, engineering, and medicine is jeopardized by the persistence of sexual harassment [5]. And since 2008, the Equal Employment Opportunity Commission (EEOC) has reported that retaliation is the most common discrimination finding in federal sector cases [14]. The most powerful deterrent of discrimination is organizational climate—the degree to which those in the organization perceive that misconduct of any form is or is not tolerated.

Institutions can take concrete steps to reduce discrimination of any sort by making systemic changes that demonstrate how seriously they take this issue and that reflect that they are listening to those who courageously speak up. This is in contrast with the policies and procedures that aim (and often ineffectively) to protect the liability of the institution but are not effective in preventing misconduct. Hierarchical power structures with strong dependencies on those at higher levels are more likely to foster and sustain sexual harassment, bullying, and other forms of discrimination. This is exacerbated in the aforementioned "rigid organizational structures," where power is concentrated in single individuals. For example, requirements that military sexual assault and harassment reports be taken up the chain of command have proven inadequate [15].

An example of all of the above is provided by the 2021 lawsuit described in [16], which was settled in favor of the complainant by a high UK court. The target is a U.S. female academic who reported sexual assault by a former UK male academic to the Human Resources of the U.S. host lab of their joint international collaborations. After no finding of fact was made against him by the U.S. lab, a long series of retaliatory events occurred, which produced significant distress on the victim that has affected all aspects of her personal and professional life.

The recommendations of the National Academy of Sciences [5] and the EEOC [14] include: 1. Creating diverse, inclusive, and respectful environments; 2. Addressing the most common form of sexual harassment: gender harassment; 3. Moving beyond legal compliance to address culture and climate; 4. Improving transparency and accountability; 5.

Diffusing the hierarchical and dependent relationship between employees and managers; 6. Providing support for the victim; 7. Striving for strong and diverse leadership; 8. Making the entire community responsible for reducing and preventing harassment and discrimination.

While the severe legal and institutional (university, laboratory, etc.) barriers that currently hamper the ability to create a just, safe, and welcoming culture in physics may take years to modify, local (collaboration, department, etc.) Community Agreements can help fill the void. These Community Agreements can delineate what behavior is appropriate and tolerated within their local community, codify those expectations, educate their members, and implement unbiased mechanisms for providing constructive feedback, with the ability to remove a member who consistently and/or egregiously harms others. Below we recommend elements for these Community Agreements and best practices for implementing them.

## 4. SCAFFOLDING ELEMENTS FOR COMMUNITY AGREEMENTS

As an example of key elements to include in a Community Agreement, we take the Core Principles and Community Guidelines created by an Ethics Task Force of the APS Division of Particles and Fields (DPF) [17]. This public document lists some of the core principles and community guidelines for DPF membership activities, and particularly for the Snowmass process itself. We note that [17] was inspired by the Princeton Physics department code of conduct and the github community guidelines, among others.

### 4.1. Core Principles and Community Guidelines

Practices of a membership community should be based on its stated underlying values, or core principles. As detailed in [17], "The core principles form the most important responsibilities of our community members." In addition to expectations from [17]: (1). Respect and support each other; and (2). Commit to constructive dialog and take initiative, are the following additional key elements:

- A mission statement that identifies the community goals.
- A signed agreement from all community members that they will abide by the norms selected by the community as a condition of membership in that community.
- A pledge to act with honesty, integrity, and courage.
- A multi-way test, such as that used in effective service organizations, such as Rotary International for instance, to check one's actions: 1. Is it the truth? 2. Is it fair to all concerned? 3. Will it build goodwill and better relationships? 4. Will it be beneficial to all concerned?

Community guidelines extend the core principles by setting expectations for participation in in-person events and virtual communication. Participation in these events and forums should be predicated on agreement to follow these guidelines as well as the APS CoC [17].

One important way to obtain consensus and buy-in of community core principles and guidelines is to train the membership, starting with leaders and others who have influence over conduct. An example of this training process already exists: conference session chair training. Because a common problem at meetings are session chairs unequipped to react to behavior violations by attendees, in 2020 both the Institute for Nuclear Materials Management and the APS DNP began session chair training for their respective meetings. The problem was also recognized by the 2022 APS Council Speaker, who advocated for adequate training for chairs at conferences and workshops [18]. It is particularly effective to use interactive role playing where session chairs are given the chance to practice interventions. The same tools and techniques can be brought to a community developing and implementing their core expectations, so that all members of the community can practice how to intervene respectfully and effectively in the moment. This peer intervention of low-level infractions can interrupt these gateway behaviors that can lead to more serious infractions.

### 4.2. Accountability and Enforceability

What is often missing in performative EDI initiatives is a clear list of increasing consequences for violating core principles and community guidelines. Without accountability for everyone's actions and transparent policies to initiate professional aftermath, attention to gender and racial equity is disingenuous and no positive change can happen. Recommendations include the following:

- Membership in the community should be conditional upon the prospective member pledging to abide by the community's Community Agreement (CA). This is one way to address a concern in [19], i.e., that any sanction taken against someone engaging in misconduct leaves community leaders who enact these sanctions open to legal action brought against them by the perpetrator. Instead, participation in the community should be predicated upon a member's agreement to uphold the goals of the community and accept the consequences if they do not.

- If the community already has bylaws, those bylaws should be amended so that they include the Community Agreement.

- Violations of the Community Agreement should be treated the same as a violation of any other element of the bylaws of, for example, an experimental collaboration. For instance, if a member of an experimental collaboration shared data with a competing collaboration without prior approval, that member can be removed from the collaboration. The standards for determining responsibility are preponderance of the evidence. The same standard should be applied to Community Agreement violations. For existing collaborations and institutions, bylaw changes may be necessary to require continuing membership to depend upon agreeing to abide by the CA. As an example of how this is handled by the APS: the APS CoC applies to every APS meeting, and some of these meetings now require consent to abide by the CoC as a condition of registration (participation). Critically, part of that agreement is the possibility that a participant will be removed from the conference without a refund should that participant be found to have violated the APS CoC.

- There need to be clear mechanisms for safe, possibly anonymous, reporting of violations of the Community Agreement, including a devoted and trained team of at least two people to whom reports can be made.

- Barriers to reporting CA concerns should be as low as possible. Ideally, every member of the community will feel safe enough to provide respectful feedback in the moment, but realistically power differentials and different responsibilities mean that the CA itself must have a mechanism for addressing concerns that cannot be addressed in the moment. To create a welcoming and inclusive culture means that we need to know when someone does not feel welcome or included. Frequent surveys to assess the climate are a better measure than waiting for complaints, but lowering the barriers to receiving complaints, supporting those who complain rather than penalizing them, will allow the community to heal most quickly.

- Some collaborations have ombuds. While the effectiveness of ombuds is still debated, if an ombuds is used they must be trained appropriately to be advocates of the process, acting independently and without pressure or fear of others. They certainly should not be used as gatekeepers. In particular, they should not be used to raise a barrier to reporting CA concerns.

- It is critical to have a mechanism for determining how to fairly and equitably respond to reports of violations of Community Agreements. Entities that are large enough may have stand-alone groups to do this and ideally have advocates available for informal mentoring of those navigating the reporting and/or enforcement process; those that are smaller may need to band together to share resources and/or appeal to an established team via the APS/DNP/DPF. Some organizations also have an EDI committee with mechanisms for investigations (see, for example, the model of the American Geophysical Union below).

- Standards of proof should be preponderance of the evidence.

- Responses to infractions should be tiered and flexible so that it is possible to respond to minor and unexpected infractions and issue warnings.

- When possible, responses to complaints should not be exclusively punitive, but should also have a restorative goal.

- When the offense is egregious or repeated, one outcome should be to inform the home institution of the defendant. Other possible consequences are listed in Section 4.

- There should be a clear procedure for removal of collaborators in case of severe violations of the CA.

- Bystander (upstander) intervention should be expected of all leaders in a community. Lack of intervention, with repeated failure to do so or failure to do so in extreme cases, should lead to sanctions for those in leadership positions. All members of the community need to be responsible for their behavior and that of others in the community. If that does not include responding in the moment, it needs to include following up appropriately, e.g., a discussion after the fact for possibly minor misunderstandings and/or reporting the incident. Having measured, tiered responses available will lower the barrier to collecting data on even minor infractions. Again, the goal is to educate and develop a community that is welcoming and inclusive.

- An appeal process should be accessible for both the initiator of the complaint and the defendant.

- The team responsible for accepting CA complaints should collect data on complaints and their outcomes. Periodic reports to the community should be made; the size of the community and the length of time between reports will determine how detailed they should be [20].

### 4.3. Other Considerations

**Limitations.** A Community Agreement that is accepted by all members of a collaboration and implemented with due process can in principle solve most problems within a collaboration, because the community has the authority to remove a problematic member if necessary. This is a necessary but not sufficient mechanism to achieve a broader physics culture that is safe, welcoming, and inclusive. CAs do not have authority beyond the community that has accepted them. A community can (and should) report a member who is behaving egregiously to their home institution, but does not have the standing to take further punitive action. For this reason, the home institution itself needs to develop effective methods for dealing with toxic behavior. But that is beyond the scope of this white paper.

**Liability Issues.** Because a collaboration is not a legal entity, a Community Agreement (CA) is not a contract, and there is always the possibility of a lawsuit from a complainant or a person sanctioned for misconduct. Having clear expectations under the CA and clear and transparent investigative procedures reduces the likelihood of success of lawsuits. At the same time, the possibility of lawsuits is an incentive against unfairly overzealous or underzealous enforcement of the CA. Organizations have sometimes hesitated to impose severe punishments, such as rescission of previously awarded honors or expulsion from the organization, for fear of civil lawsuits, especially when the finding of misconduct is based on confidential data. This exposure can be largely eliminated by a two-stage process suggested by Leonid Levin at the 2022 annual meeting of the National Academy of Sciences. In the first stage, the adjudicating committee, acting on partly confidential data, would draw up a redacted summary of the misconduct they had found, along with a proposed punishment. The accused member would then be given the option of accepting the punishment or appealing it to a vote of the membership. While a disgruntled former member might successfully argue that a secretive committee had defamed them, it would be hard to argue that the whole organization had done so. Being thus better able to expeditiously and safely punish severe misconduct, the organization would be better able to avoid lawsuits of the opposite kind, by complainants who felt their complaints had been ignored and their rights violated by an organization too fearful of being sued for overzealous enforcement.

**Transparency.** Since inequity, or injustice, can only be sustained with the lack of transparency, it is of paramount importance that those charged with investigating CoC violations track all reported violations and keep detailed confidential records of their handling of each complaint. The tradeoff between transparency and confidentiality is delicate. It needs to be navigated while keeping in mind that too much transparency early on discourages frank reporting and investigation, while too little transparency at the end leaves the community wondering if complaints have been properly investigated or even whether investigators are corruptly excusing friends and punishing enemies. In the intermediate term appropriately anonymized or redacted data can be made publicly available, as done in [17]. To maximize transparency, the rules for who can examine still-confidential data, and when, should be publicly known and hard to change. A possible final incentive for ethical conduct by investigators would be to require the complete confidential record of their deliberations to be made public after, say 50 or 75 years. In other words, caring about one's legacy should be an informal prerequisite for serving as a CoC investigator.

**Ethics/EDI Committees.** Once the elements of a Community Agreement have been defined and accepted by the community members, the individuals in charge of it are tasked with the challenging responsibility of implementing the rules as objectively as possible. Some organizations have an EDI Office/Committee and some others have an Ethics Committee. Ethics Committees are sometimes charged with oversight of Ethics Policy content and considerations, including potentially enforcements and investigation recommendations. The EDI committees are typically charged with advancing an organization's total inclusive science culture, and assuring attention to addressing EDI programming needs. In the example of the American Geophysical Union AGU, both an EDI Office led by an EDI Chief and an Ethics Committee of volunteers are present. The details will depend upon the size of the organization, available resources, and possibly existing organizational structures. For purposes of this discussion we will use "EDI Committee" (EDIC) to refer to the group of people who accept complaints of potential CA violations and determine how to respond to them.

**Implementation Bias.** For Community Agreements to positively impact climate in the community, they have to be put in practice in the most constructive and transparent way. All individuals possess bias, including members of EDI Committees. People who serve on EDICs should undergo, at least annually, training where they practice on case studies and learn how to identify their own biases and mitigate them. The APS has used a professional trainer to create the DNP Allies program [21] that could perform this training, but other options are available. Not only do humans have biases with regard to gender, race, disability, etc., but they can be more or less subject to inappropriate pressure within the organizational structure or beyond it, whether consciously or not. Issues that can cloud a fair and objective investigation of a complaint, as well as implementation of effective remedies include: (1). Fear of retaliation or impact on one's career, particularly a concern if one were to find an influential person responsible for misbehavior. This is of relevance for instance when younger, earlier career individuals are in charge of making ethical decisions on cases involving more established or well-known defendants. (2). The impulse to please, or a conciliatory attitude to keep superficially good relationships. This can lead to the desire to "smooth over" concerning behavior and consider misbehavior as "not so bad." It might occur if offenders are personally known to EDIC members, or if they are well-connected to an exclusive network (e.g., the "old boy network"), and might afford more freedom in words and behavior within their exclusive group. Such considerations can also influence any oversight groups (e.g., Boards of Directors) a community may have, who might want to veto the findings and/or recommended remedies of the EDIC. (3). A sense that if someone is important enough and/or accomplished enough, behavioral norms don't apply to them. Physics in particular has seen the phenomenon that if a scientist is brilliant enough, their poor behavior may be excused as the price the community is willing to bear to have this brilliance among them. (4). The instinct to "save face," i.e., do all that is in one's power to avoid responsibility for one's actions when performing poorly. This can be exhibited through an EDIC reaching the wrong decision on a case and then blaming the victim to justify their poor response. Although protecting one's reputation may be a major concern for all, it is a driving influence for those who lack competence and humility [22].

**Checks and Balances.** Every complaint received by the EDIC can inform whether the CA elements and processes need to be updated or modified. Annual activities should include: (1). The CA and policies should be reviewed in the context of complaints received and how they were handled. (2). The community culture should be measured, at least by administering an anonymous survey and possibly by focus groups. If the community is too small for anonymity to be assured, that survey can be administered and reviewed by the APS and/or another entity agreed upon by the community. (3). A report with the appropriate level of confidentiality should be made available to community members so that they can help determine whether the CA needs to be modified. (4). The community should discuss and reaffirm their commitment to the (possibly modified) CA, and discuss case studies that will help the members understand the importance of the CA to the functioning of the community.

## 5. THE AMERICAN GEOPHYSICAL UNION (AGU), A ROLE MODEL

The American Geophysical Union (AGU) Scientific Integrity and Professional Ethics Policy [23] is an example of a scientific professional organization's policy with explicit expected standards for behavior and also effective procedures for holding accountable those who violate those standards. It was established on the premise that scientific integrity and ethics are fundamental to scientific advancement and science cannot flourish without the respectful and equitable treatment of all those engaged in the scientific community. To address ongoing issues within the scientific community, a majority of which do not rise to the level of legal actions, yet have profound impact on individuals lives and career, the AGU policy has established its own definition of scientific misconduct. For instance, AGU expanded the "Scientific Misconduct" definition to include misconduct towards others. They have also expanded the definition of harassment and discrimination in addition to any legal definition, and have added bullying, which is not illegal by U.S. law, to the list of unacceptable behavior.

The AGU, an organization with ~62,000 members, uses the following procedures for handling complaints. The Executive Vice President for EDI at AGU and the chair of the AGU Ethics committee (composed of AGU members) decide together which of the received complaints warrant a full investigation, in coordination with the AGU President and with the AGU CEO. For each investigation, the Executive Vice President for EDI chairs an ad-hoc committee with the following composition: 1. An attorney as external consultant; 2. The chair of the AGU Ethics committee (composed of AGU members); 3. Two ethics specialists, who are experts in their field, such as former EDI chiefs or Ethics Professors who donate their time; 4. One other person of competence in the case.

The ad-hoc committee investigates whether there is compelling evidence for unethical and/or unacceptable behavior. They are not searching for violations of federal or state statutes; this is not a court of law; they are a community deciding whether their community agreements have been violated. The goal is to discourage scientific misconduct within the ampler definition set by AGU, not to decide on issues with legal implications. Importantly, institutional findings are not always required. The expert composition of the committee ensures that each case is handled with the greatest possible objectivity. The total effort for each investigation is usually five meetings of two hours each with the various parties [24].

With clear policies and procedures in place, even a non-profit organization has the authority to change the status of its members. This gives the community opportunities, limitations, and obligations significantly different from those of judicial or governmental bodies. Using these the AGU is able to encompass protection of their members for both of the following cases:

- The member is negatively affecting AGU programs and/or AGU members.
- The member is causing reputational damage to the organization. This latter condition is a way to discipline bad actors at other institutions when such bad actors are AGU members – especially those holding high AGU honors or elected/appointed volunteer positions.

Examples of increasing sanctions available to a community include 1. Written warning; 2. Notification sent to respondent's home institution; 3. Removal from volunteer positions; 4. Suspension of membership; 5. Permanent expulsion; 6. Denial or revocation of honors and awards; 7. Notification sent to members of incident; 8. Public statement regarding misconduct.

The lower levels 1-4 offer the community a timely way to censure and distance themselves from members who cause reputational damage to the organization, even if, for personnel confidentiality or whistleblower protection reasons, the misconduct can't be publicly revealed. Some of these measures, including conditions for denial or revocation of honors and awards and removal from official positions, have already been established within the American Physical Society, and are overseen by the APS Ethics Committee. The APS CEO Jon Bagger is currently looking to establish additional procedures at the APS.

The exemplary standards of behavior developed and established at the AGU should inform more empathic, modern and homogenous procedures for the Human Resources (HR) and Legal Counsel offices in any scientific organization funded by the taxpayers. For instance, it is well-known that deeply grounded cultures of nepotism and cronyism, as those found in some national labs [25], establish strong authority in the hands of a few, as well as extended unethical human treatment of many, consequently creating what some perceive as a toxic climate at several of these labs. In order to take a measure of climate, the National Academies recommend doing a third-party anonymous survey as, indeed, waiting for complaints is an unreliable way of measuring climate [5].

A textbook consequence of ineffective handling of a discrimination complaint at a U.S. national lab was made public in September 2021. A lab settled with the EEOC on a claim that it had "denied promotion to a female engineer after she complained about sex-based discrimination." Besides the monetary relief, the settlement included that the lab display a posting regarding the resolution of the lawsuit to its employees, provide training on retaliation under Title VII to its employees, and make regular reports to the EEOC regarding its compliance with the decree [26]. All national labs would greatly benefit by applying the same increased standards of behavior as the AGU, which have proven extremely effective in improving climate in the Earth and space science communities.

Scientists also have ethical obligations to the public. Ethics, normally considered a branch of philosophy, draws on fields of social science where authors routinely supplement the methods common to all sciences by trying to analyze their own unconscious preconceptions, and explain the resulting imperfect understanding in a "positionality statement" for the benefit of readers. Such introspection is rarely practiced by physicists, whose objects of study lack emotions, needs, rights, etc.; but in both cases the goal is to maximize the transparency and reconstructability of what one is doing. This is one of several ways natural and social scientists can learn from each other and improve science's reputation for truthfulness and humility [15]. Accordingly, it would be wise for designers and implementers of Community Agreements to approach their task with similar humility, mindful of their own fallibility and the fact that notions of ethical conduct can change significantly over the course of a human lifetime and are likely to evolve further in the future.

## 6. CONCLUSIONS

A considerable effort has been devoted by various groups to developing Community Guidelines. When accompanied by accountability and enforcement processes, they develop into effective Community Agreements. Their beneficial impact for the current community and the positive investment for Physics' future generations make these a critical part of our culture going forward. All physics collaborations, however large or small, should have a Community Agreement in place. For those communities that already have bylaws, Community Agreements could become part of the bylaws. However, for our entire physics community to be safe and welcoming will require more than local collaboration/department/etc. Community Agreements. Ultimately, institutions, laboratories, universities, professional societies, and funding agencies will need to commit the resources and undertake the responsibility of providing and implementing effective strategies to achieve the broader communities' goals. Community Agreements are something that each collaboration can do *now* until further help is available. An institution's leadership or a collaboration's spokespersons have to support EDI committees in full, and help them integrate as seamlessly as possible within the organization. If they become a cohesive, active, and positive part of the organization, established EDI committees could be in charge of the Community Guidelines and of their enforcement. Community Agreements should be reviewed and updated on a yearly basis, implementing criteria of accountability and transparency as needed.

Like any other formal mechanism, Community Agreements, no matter how well designed, are susceptible to being ignored or gamed, except in a culture with sufficient mutual trust and determination to implement them honestly. Results from social sciences can help cultivate this culture. Indeed, while the statistics of abuse in our field are disturbing, each of these statistics represents a human who has been injured. Not by strangers but by collaborators, colleagues, and peers in physics. Too many physicists think that if they are not personally abused or affected, it is not their problem what happens to those who are. Some physicists push back on the concept of Community Agreements because it is "too hard" to hold abusers accountable, or because they have already tried to build effective Community Agreements and have failed. Some fear that "careers would be upended" and the reputation of the community would suffer if complaints of harassment, retaliation, or defamation were made. If you are one of those physicists, consider the impact on the victims' careers, and the reputational damage done to communities when bad behavior is allowed unchecked. Some conservative institutions want to avoid Community Agreements – or at least their enforcement – for fear of lawsuits. Those institutions might want to think about their vulnerability to lawsuits brought by those who reported bad behavior but were ignored. Otherwise, we will continue the current status where (1). Many members of marginalized groups feel unsafe in their professional environment, including at conferences. This is clearly destructive both to individuals and to the scientific endeavor; (2). Many members of the community experience health damages as described in [27], with some targets of more egregious offenses developing Post Traumatic Stress Disorder (PTSD); (3). Victims and targets suffer further damage to careers and to their reputations when broken HR systems use inappropriate standards and determine that nobody is responsible for their experience; (4). Perpetrators are empowered to retaliate once the victims' complaints are dismissed; (5). The field of physics continues to lose the skills of people not willing to sacrifice their health and dignity to work in a culture they perceive as toxic. As a scientist, please consider it part of your job to contribute to a safe and welcoming culture in physics. To quote Tim Hallman, Associate Director for Nuclear Physics at the DOE: "This has to stop … The only thing it takes for bad-behavior defeating DEI goals to continue is for people of good conscience and integrity to do nothing."

## Acknowledgments

The authors are grateful to Billy Williams, AGU, Executive Vice President, EDI; Prof. William A. Barletta, Adjunct Professor, MIT; Prof. Matthew G. Hannah, Chair in Cultural Geography, University of Bayreuth, Germany, Jeremy Wolcott, Tufts University, and Michael Troxel, Duke University, for their valuable feedback.